\documentclass[twocolumn, times]{aastex631}
\usepackage{graphicx}
\usepackage{amsmath}
\usepackage{amssymb}
\usepackage{natbib}
\usepackage{hyperref}
\usepackage[all]{hypcap}
\usepackage{xcolor}
\usepackage{soul}
\usepackage{mathtools}
\usepackage{bm}

\hypersetup{
    colorlinks=true,
    linkcolor=blue,
    filecolor=magenta,
    urlcolor=blue,
}

\newcommand{\fD}{f_\mathrm{D}}
\newcommand{\vect}[1]{\boldsymbol{#1}}

\begin{document}

\title{Using Giant Pulses to Measure the Impulse Response of the Interstellar Medium}

\author[0000-0002-6317-3190]{Nikhil Mahajan}
\affil{Department of Astronomy and Astrophysics, University of Toronto, 50 St. George Street, Toronto, ON M5S 3H4, Canada}

\author[0000-0002-5830-8505]{Marten H. van Kerkwijk}
\affil{Department of Astronomy and Astrophysics, University of Toronto, 50 St. George Street, Toronto, ON M5S 3H4, Canada}

\correspondingauthor{Nikhil Mahajan}
\email{mahajan@astro.utoronto.ca}

\begin{abstract}
Giant pulses emitted by PSR B1937+21 are bright, intrinsically impulsive bursts.
Thus, the observed signal from a giant pulse is a noisy but direct measurement of the impulse response from the ionized interstellar medium.
We use this fact to detect 13,025 giant pulses directly in the baseband data of two observations of PSR B1937+21.
Using the giant pulse signals, we model the time-varying impulse response with a sparse approximation method, in which the time dependence at each delay is decomposed in Fourier components, thus constructing a wavefield as a function of delay and differential Doppler shift.
We find that the resulting wavefield has the expected parabolic shape, with several diffuse structures within it, suggesting the presence of multiple scattering locations along the line of sight.
We also detect an echo at a delay of about 2.4 ms, over 1.5 times the rotation period of the pulsar, which between the two observations moves along the trajectory expected from geometry.
The structures in the wavefield are insufficiently sparse to produce a complete model of the system, and hence the model is not predictive across gaps larger than about the scintillation time.
Nevertheless, within its range, it reproduces about 75\% of the power of the impulse response, a fraction limited mostly by the signal-to-noise ratio of the observations.
Furthermore, we show that by deconvolution, using the model impulse response, we can successfully recover the intrinsic pulsar emission from the observed signal.
\end{abstract}


\section{Introduction}

Radio signals emitted by pulsars propagate through the ionized interstellar medium (ISM) and are distorted by a variety of frequency-dependent effects such as dispersion, birefringence, and scintillation due to multi-path scattering \citep{Rickett1990}. While these distortions make it difficult to observe the intrinsic pulsar emission at low radio frequencies, they also contain a wealth of information about the structure of the ISM along the line of sight to the pulsar.

If the pulsar emission can be coherently separated from propagation effects in the ISM, both the intrinsic radio emissions of pulsars and the structure of the ISM can be better studied. For example, coherent dedispersion \citep{Hankins1975} is used to remove the $\nu^{-2}$ dispersive effects of the ionized ISM, by applying an inverse filter to raw voltage (baseband) data, for a known dispersion measure (DM). Circular birefringence caused by Faraday rotation can also be corrected in a similar manner using a rotation measure (RM).
There is, however, no similar general technique for inverting the multi-path scattering effects responsible for scintillation.

Traditionally, pulsar scintillation is studied using dynamic spectra. Recovering the impulse response of the ISM from a dynamic spectrum requires solving an ill-posed phase retrieval problem, since generally only the amplitude information is kept in the process of creating a dynamic spectrum.
Thus, approaches to this problem so far \citep{Walker2008, Baker2022} usually impose a sparsity constraint in order to tackle it. \cite{Oslowski2022} show that these techniques fail when the true impulse response is dense.

Alternatively, cyclic spectroscopy \citep{Demorest2011, Walker2013} is a technique that exploits the cyclic nature of pulsar emission, to generate ``cyclic spectra'', which preserve some of the phase information that would traditionally be lost in dynamic spectra.
These cyclic spectra can then be used to decouple the impulse response of the ISM from the intrinsic pulse profile, as demonstrated by \cite{Walker2013}.
However, this technique assumes the observed signal is cyclostationary (i.e., the pulse profile is stable over the integration time).
Transient emission phenomena such as nulling, mode changing, or giant pulse emission would violate this assumption, and could affect the validity of a cyclic spectrum.

In this work, we present a technique for using raw baseband data of bright and impulsive giant pulses as direct measurements of the ISM's impulse response, and apply it to observations of \object{PSR B1937+21}.
While this technique does not depend on a priori knowledge about the source or assumptions about the structure of the impulse response, it does require the source to produce bright impulsive bursts at a relatively high rate.
Using the fact that nearby impulses are imprinted with the same response, we find thousands of giant pulses, which we then use to model the time-varying impulse response of the ISM.
We discuss how this modeled impulse can help us to study the structures in the ISM which cause scintillation, or recover the intrinsic pulsar emission from the observed signal to better understand the radio emission mechanism in pulsars.
\section{Background}

For a point-like source such as a pulsar, propagation effects through the ionized ISM can be treated as linear and time-invariant over short timescales. When the pulsar emits a signal $x(t)$, we observe
\begin{equation}
    y(t) = (h \ast x)(t) + \epsilon
\end{equation}
where $\ast$ denotes a convolution, $h$ is the impulse response function (IRF), and $\epsilon$ is the noise term. This IRF includes dispersion, scattering, and scintillation effects due to multi-path propagation, as well as any instrumental effects.
Due to the relative motion of the pulsar, the Earth, and the structures in the ISM, the IRF evolves with a characteristic timescale usually called the scintillation timescale, $t_\mathrm{scint}$.
So, the effects of the ISM can be properly characterized by $h(\tau, t)$, the instantaneous impulse response at time $t$, where $\tau$ is the relative delay (also known as the lag).
For most sources, the timescales are well separated: typical scintillation timescales are of order a few minutes, while maximum lags at which the IRF still has power are of order a millisecond.

If a source emits an impulse at time $t_0$, such that $x(t) = a_0 \, \delta(t - t_0)$, with $a_0$ a complex-valued amplitude, then we observe $y(t) = a_0 \, h(t - t_0, t_0) + \epsilon$.
We can shift the observed signal in time to acquire
\begin{equation}\label{eq:g}
    g(\tau) = y(\tau + t_0) = a_0 \, h(\tau, t_0) + \epsilon,
\end{equation}
a noisy measurement of the instantaneous IRF at time $t_0$ scaled by a complex amplitude which is a property of the emitted impulse.
The impulse response must naturally be causal such that $h(\tau, t) = 0 \, \forall \tau < 0$.

Pulsars that emit giant pulses can be used for such measurements of the IRF, since giant pulses are so short to be unresolved (in relatively narrow frequency bands).
In order to properly measure and model $h(\tau, t)$, the pulsar must emit sufficiently bright pulses at a sufficiently high rate such that multiple good measurements of the IRF can be made within the scintillation time.
If impulses are emitted too sparsely, then it may not be possible to recover information about the more rapid variations in $h(\tau, t)$.
The bright giant-pulse emitter PSR B1937+21, which produces thousands of impulsive giant pulses per hour, is suitable for fully modeling $h(\tau, t)$.

\subsection{Conjugate Wavefield}

Pulsars emit giant pulses irregularly, which means that any model of the IRF, $h(\tau, t)$, requires an interpolation scheme.
A convenient one is to write $h(\tau, t)$ as a Fourier series with terms regularly spaced in frequency.
The Fourier conjugate of $t$ is $-\fD$, where $\fD$ is the differential Doppler shift relative to the line of sight (following astronomical convention of a Doppler shift being positive as the source is moving away from the observer).
In the usual physical picture where the delay $\tau$ reflects the extra path length introduced by scattering off a structure some distance away from the line of sight, one has $\fD = \dot{\tau} \nu$.

We define a ``conjugate wavefield'',
\begin{equation}\label{eq:cw}
    W(\tau, \fD) \coloneqq {{\cal F}_t}^{-1} \left[ h(\tau, t) \right],
\end{equation}
where ${{\cal F}_t}^{-1}$ is the inverse Fourier transform along $t$.
With this, the IRF at any time $t$ can be determined by
\begin{equation}\label{eq:hpred}
    h(\tau, t) = \sum_{\fD} W(\tau, \fD) \, e^{-2 \pi i \fD t}.
\end{equation}

\subsection{Relation to Dynamic and Secondary Spectra}

In pulsar scintillation studies, a dynamic spectrum, $I(\nu, t)$, is usually measured, and often the so-called ``secondary spectrum'', $S(\tau, \fD)=|{\cal F}_{\nu, t}[I(\nu, t)]|^2$ (where ${\cal F}$ denotes a Fourier transform, here along frequency $\nu$ and time $t$), is calculated, because in $\tau, \fD$ space the underlying structure of the scattering screen is much more apparent.

The dynamic spectrum is related to the time-varying IRF via
\begin{equation}
    I(\nu, t) = \left| {{\cal F}_\tau} \left[ h(\tau, t) \right] \right|^2
\end{equation}
(where we assume that the possible (slow) variation of the pulsar emission strength with frequency has been removed).
Hence, the secondary spectrum is related to the auto-correlation of conjugate wavefield via $S(\tau, \fD) = \left| W \star W \right|^2$ (where $\star$ denotes cross-correlation).

\section{Observations} \label{sec:obs}

We observed PSR B1937+21, a $1.56$ ms pulsar, with the 327 MHz Gregorian receiver at the Arecibo Observatory for $2$ hours on 2021 May 7 (MJD $58245$), and $30$ minutes on 2021 May 29 (MJD $58298$).
Using the Puerto Rico Ultimate Pulsar Processing Instrument (PUPPI) in raw baseband mode, we recorded 32 contiguous $3.125$ MHz bands of dual-polarization baseband data (i.e., a sample spacing of $320$ ns).
A polyphase filter is applied in the PUPPI backend which reduces spectral leakage between bands.
We use the \textsc{Baseband} \citep{baseband} and \textsc{Pulsarbat} \citep{pulsarbat} software packages to read and process the raw baseband data.
For our analysis, we only use bands in the range of $297.3125$ to $356.6875{\rm\;MHz}$ as the remaining bands fall outside the analog bandpass filter and have diminished signal strength.
This leaves us with 19 usable frequency bands, with a total bandwidth of $59.375$ MHz centered on $327$ MHz.

We pre-process the data by normalizing the passband to correct for the effects of the polyphase filter and any bright narrow-band channels that may occur due to radio frequency interference (RFI).
For both observations, we experience insignificant amounts of RFI and so we apply no further RFI mitigation.
The data are then coherently dedispersed using a DM of $71.0201{\rm\,pc\,cm^{-3}}$ on MJD $58245$, and $71.0169{\rm\,pc\,cm^{-3}}$ on MJD $58298$, with the DM values inferred from giant pulses found in the dataset.
For the technique presented in this paper, it is not important for the data to be perfectly de-dispersed, since any excess dispersion will be captured as part of the ISM's impulse response.

We correct for relative time delays between the left and right circular polarizations of $7.85 \, \mathrm{ns}$ (MJD $58245$) and $37.85 \, \mathrm{ns}$ (MJD $58298$)\footnote{We think that the relative difference of exactly $30$ ns is caused by the polarization pipelines in the instrument backend being out of sync by 3 samples when processing the raw 100 MHz signal stream.}.
These delays are dominated by instrumental effects, but also include a contribution from circular birefringence in the ISM due to magnetic fields, also known as Faraday rotation, characterized by a rotation measure (RM) of $\sim \! 7$ to $8{\rm\,rad\,m^{-2}}$ \citep{Yan2011, Dai2015, Wahl2022}.
\section{Giant pulse search} \label{sec:gp_search}

Giant pulses, being bright narrow bursts, are usually easy to detect.
However, to be useful as measurements of the ISM's impulse response, we need to align their observed signals to each other in $\tau$ (see Equation~\ref{eq:g}) to well within a sample.
We achieve this by precisely measuring the differences in time of emission between neighbouring giant pulses.
Note that we use the term ``giant pulse'' to refer to any sufficiently bright impulse detected by our technique, without a specific notion of what causes giant pulse emission.

We split up our observation into blocks much shorter than $t_\mathrm{scint}$ such that the IRF can be taken to be essentially time-invariant across a few adjacent blocks.
We then use an iterative procedure where giant pulses found in one block are used to find giant pulses in adjacent blocks and so on until all blocks have been searched.

\begin{figure}
  \centering
  \includegraphics[width=0.45\textwidth,trim=0 0 0 0,clip]{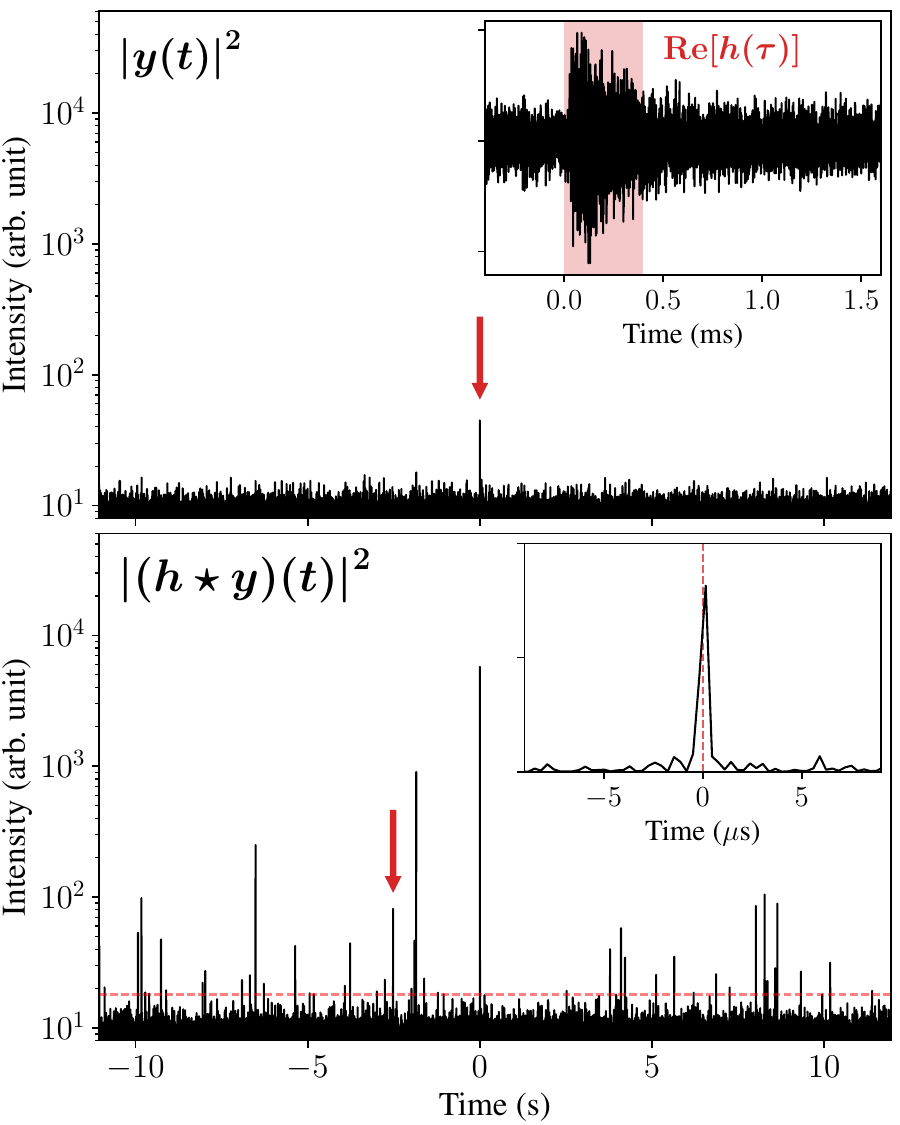}
  \caption{
    Giant pulse search.
    {\it Top:\/} Squared-modulus of the observed signal.
    The reference pulse is marked with an arrow.
    The inset shows the real part of the complex-valued voltages of the reference pulse, with the $0.4$ ms span used in the cross-correlation to find further pulses highlighted in pink.
    {\it Bottom:\/} Signal cross-correlated with the reference pulse, revealing many other  impulses.
    The dashed horizontal line denotes the detection threshold we use.
    The inset is a zoom-in around the impulse marked with the arrow, with the dashed vertical line indicating the impulse's computed location.
  }\label{fig:descatter}
\end{figure}

In one cycle of this iterative loop, the goal is to find impulsive giant pulses in a given block centered on some time $t'$.
We have $y(t)$, the coherently dedispersed observed signal (processed as described in Section~\ref{sec:obs}), and a set of previously detected giant pulses (in neighbouring blocks) of the form,
\begin{equation}
g_j(\tau) = a_j \, h(\tau, t_j) + \epsilon_j,
\end{equation}
such that $|t_j - t'| < \Delta t_\mathrm{max}$, and $a_j$ are the a priori unknown complex amplitudes of the pulses.
The chosen $\Delta t_\mathrm{max}$ must be much smaller than the scintillation timescale, $t_\mathrm{scint}$, such that we can safely treat these giant pulses as approximations of $h(\tau, t')$.
In our data, we find that $t_\mathrm{scint} \approx 70 \, \mathrm{s}$, and we choose $\Delta t_\mathrm{max} = 20 \, \mathrm{s}$.
Since we expect the signal-to-noise ratio of the IRF to drop exponentially with delay, and since the parts at high delay vary faster (since generically these deviate more from the direct line of sight), we truncate the giant pulses.
We chose $\tau_\mathrm{max} = 0.4{\rm\,ms}$, which captures approximately $90 \%$ of the power in the IRF.

Since the observed signals are discretely sampled, we can form a matrix $\vect{G}$ with elements $\vect{G}_{j,k} = g_j(\tau_k)$ such that
\begin{equation} \label{eq:svd}
    \vect{G} \approx \vect{a} \vect{h}^\intercal,
\end{equation}
where $\vect{a}$ is the vector of amplitudes $a_i$, and $\vect{h}$ is the discretely-sampled impulse response $h(\tau, t')$. Since $\vect{a} \vect{h}^\intercal$ is a rank-1 approximation of $\vect{G}$, we can use singular value decomposition (SVD) to estimate $\vect{h}$ (to within a complex-valued amplitude) as the first right singular vector of $\vect{G}$ (by the Eckart--Young--Mirsky theorem).

If we cross-correlate this estimate of $h(\tau, t')$ with $y(t)$, giant pulses which are intrinsically impulsive will show up as bright impulses,
\begin{equation} \label{eq:cc}
    z(t) = (h \star y)(t) = \sum_j c_j \, \delta(t - t_j) + \epsilon,
\end{equation}
where $\epsilon$ is the noise term which includes all other contributions to the cross-correlation, $t_j$ is the location of the $j$-th impulse in time, and $|c_j|^2 / \langle |\epsilon|^2 \rangle$ represents the effective signal-to-noise ratio of the giant pulse detection.
We describe our giant pulse detection criteria in Section~\ref{ss:detect}.

For detections considered significant enough, we precisely measure $t_j$ using an impulse estimation technique described in Appendix~\ref{app:impulse}.
With that, we can extract time-shifted snippets of the observed signal, $g_j(\tau) = y(\tau + t_j)$ to get noisy measurements of the IRF at $t_j$ with unknown amplitudes $a_j$ (Equation~\ref{eq:g}).
These are then added to our growing set of detected giant pulses and used to find more giant pulses in other cycles of the iterative loop.
To start the process, we begin by manually finding a very bright and preferably broadband giant pulse which serves as a reference for the giant pulse search (specifically, a reference for $\tau = 0$).
The iterative loop is started with the block containing this first giant pulse. We illustrate this in Figure~\ref{fig:descatter}.

\subsection{Giant Pulse Detection Criteria} \label{ss:detect}

To ensure we get no false positives, we set strict detection criteria when finding the impulses in $z(t)$ from Equation~\ref{eq:cc}.
The noise term is well described by complex-valued additive Gaussian white noise, and thus $|\epsilon|^2 \sim \chi_2^2$ (the chi-squared distribution with 2 degrees of freedom).
For detecting peaks, we use a minimum signal-to-noise threshold of $|c_j|^2 / \langle |\epsilon|^2 \rangle > 18$.
We can measure the location of an impulse with a standard deviation of $\sigma_{t_j} \lesssim 30{\rm\,ns}$ or $\sim\!0.09$ samples (see Appendix~\ref{app:impulse}).

This signal-to-noise threshold gives us a false positive rate of $10^{-7.3}$.
However, an impulse is only accepted if it is detected in a minimum of 3 signal streams out of 38 (from 19 frequency bands and 2 polarizations).
The expected number of false positives across all our data is thus much less than one.
By requiring that an impulse must be detected in at least two frequency bands, we avoid accidentally detecting any bright narrowband RFI bursts.

We also require that all detections of the same impulse are no more than $200$ ns apart from each other in time across the various signal streams.
We use the median time of all detections to estimate $t_j$, which lowers the error to $\sigma_{t_j} \lesssim 20{\rm\,ns}$ for the weakest pulses while being robust to outliers.
Finally, we filter out detected pulses which are closer than $0.6$ ms to each other since these signals would essentially contain a mixture of two differently-delayed copies of the impulse response.

With these criteria, we detect 9627 and 3398 giant pulses on MJDs 58245 and 58298, respectively.
This corresponds to detection rates of $1.5$ and $2.1{\rm\,s^{-1}}$, respectively.
The baseband snippets for these 13,025 detected giant pulses are made available as a dataset on Zenodo at \dataset[doi:10.5281/zenodo.7901384]{\doi{10.5281/zenodo.7901384}}.
\section{Modelling the Wavefield} \label{sec:wavefield}

From the giant pulse search, we have thousands of noisy measurements of the IRF, $g_i(\tau)$.
Using Equations~\ref{eq:g} and \ref{eq:hpred}, we can form a system of equations that needs to be solved for the wavefield,
\begin{equation}\label{eq:lin-sys}
    g_j(\tau) = a_j \, \sum_{\fD} W(\tau, \fD) \, e^{-2 \pi i \fD t_j} + \epsilon_j.
\end{equation}
For every $\tau$, we have a linear system of the form $\vect{y} = \vect{A} \vect{w} + \epsilon$ where $\vect{y}$ and $\vect{w}$ are the data and wavefield vectors, respectively, at a particular $\tau$, and $\vect{A}$ is the coefficient matrix with elements given by
\begin{equation}
  A_{jk} = a_j \exp(-2\pi i {\fD}_k t_j).
\end{equation}
The linear systems for different $\tau$ are linked only through the amplitudes $a_j$, which are not known a priori.
We describe our method for estimating $a_j$ in Section~\ref{ss:amplitude}, after discussing how we solve for $W$ for known $a_j$.

\subsection{Orthogonal Matching Pursuit} \label{ss:omp}

Due to the low signal-to-noise ratio in the dataset, ordinary least squares without regularization gives poor results.
Since we expect the wavefield to be at least globally sparse (i.e., the observed scattered signal occupies a small portion of the $(\tau, \fD)$ space), a sparse approximation technique should be appropriate.
We use Orthogonal Matching Pursuit (OMP), an iterative greedy algorithm for approximating sparse signals \citep{Tropp2004a, Cai2011}, with our choice determined mainly by it being computationally fast and easy to implement relative to most other sparse approximation or signal recovery techniques.

In each step of OMP, we select the $\fD$ column of $\vect{A}$ which is most correlated with the current residual vector and add it to a set of selected columns.
Hence, we iteratively approximate the signal vector as $\hat{\vect{w}}_s = (\vect{A}^*_s \vect{A}_s)^{-1} \vect{A}^*_s \vect{y}$, where $\vect{A}_s$ is the submatrix of $\vect{A}$ that only contains the selected columns, then calculate a new residual vector $\vect{r} = y - \vect{A}_s \hat{\vect{w}}_s$, and make a new approximation.
Here, $^*$ denotes a conjugate transpose.
This is repeated until a stopping criterion is reached.
The algorithm returns the best approximation of the signal vector, $\hat{\vect{w}}$, which is non-zero only for elements corresponding to $\hat{\vect{w}}_s$.

We use a stopping criterion of $\| \vect{A}^* \vect{r} \|_\infty^2 > \gamma \| \vect{a} \|^2$, where $\vect{a}$ is the vector of amplitudes. This essentially stops the algorithm when our residuals no longer strongly correlate with any $\fD$ column in $\vect{A}$.
According to \cite{Cai2011}, this stopping criterion is effective when the noise is Gaussian.
If all the signal components in the actual wavefield were very bright, then we would be able to perfectly recover the wavefield using OMP under this stopping criterion.
However, in natural signals, the magnitudes of sorted components decay as a power law and there is usually no clear boundary between signal and noise.
Thus, the choice of $\gamma$ presents a trade-off between denoising and accuracy. The variance of noise that ends up in the modeled wavefield decreases exponentially with increasing $\gamma$.

\subsection{Warm start} \label{ss:amplitude}

Initially, both the amplitudes of the detected giant pulses $a_j$ and the wavefield $W(\tau, \fD)$ are unknown to us.
However, from Equation~\ref{eq:lin-sys}, if the wavefield is known, we can solve for the amplitudes, and vice-versa.
We use this idea to ``warm start'' our optimization routine.

Given a giant pulse, $\vect{g_j} = a_j \vect{h_j} + \epsilon$, we can estimate $|a_j|$ via
\begin{equation}\label{eq:ampsolve}
    |a_j|^2 \approx \frac{\| \vect{g}_j \|^2 - N \sigma^2}{\| \vect{h}_j \|^2},
\end{equation}
where $N$ is the length of the data vector and $\sigma^2$ is the variance of the noise term, $\epsilon$, which is assumed to be additive Gaussian white noise.
The giant pulse needs to have a high signal-to-noise ratio to ensure that $\| \vect{g}_j \|^2 > N \sigma^2$.
Without flux calibration, we cannot determine the total integrated intensity in the impulse response (sometimes called the magnification).
So, we assume that $\| \vect{h} \|^2 = 1$ (which is reasonable for our case, since the density in $\fD$ and $\tau$ of our solutions implies that the radiation arrives to us via many different paths).

Given two giant pulses, $\vect{g_j}$ and $\vect{g_k}$, close to each other in time such that they approximate the same impulse response $\vect{h}$, we can also estimate the relative phase difference between their amplitudes via,
\begin{equation}
    a_j a_k^* \approx \frac{\vect{g_j} \, \vect{g_k^*}}{\| \vect{h} \|^2}.
\end{equation}
This can be iteratively applied to determine the phases of all giant pulses relative to some reference pulse. The absolute phase is arbitrary and cannot be determined.

Due to the low signal-to-noise ratio in our giant pulse dataset, the amplitudes cannot be reliably estimated directly.
Instead, we use overlapping $20 \, \mathrm{s}$ intervals with a step size of $10 \, \mathrm{s}$ and construct higher signal-to-noise approximations $\vect{h}_{{\rm SVD},i}$ of the IRF at the center of each interval $i$ using the SVD-based technique from Equation~\ref{eq:svd}.
The amplitudes $a_{{\rm SVD},i}$ for these approximations are then easily estimated.
We use overlapping intervals to get more reliable estimates for the relative phases between adjacent approximations.

With the amplitudes $a_{{\rm SVD},i}$, we then solve for an approximation of the wavefield using Orthogonal Matching Pursuit as described in Section~\ref{ss:omp}.
This wavefield approximation differs from the actual wavefield since combining information from giant pulses in an interval essentially applies a sinc filter on the wavefield.
Furthermore, as the times at which giant pulses occur are essentially random, the times of the approximated IRF have some jitter around the mid-points of the intervals, based on how the giant pulses were weighted in the SVD.
For instance, a very bright giant pulse at the edge of an interval will dominate the approximation, leading to a jitter of almost half the interval width.

We can, however, use this approximated wavefield to estimate the amplitudes of the individual giant pulses using Equation~\ref{eq:lin-sys}. Then we use these amplitudes again to solve for a better model of the wavefield using the giant pulse data directly.

\subsection{Wavefield Solutions}

\begin{figure*}
  \centering
  \includegraphics[width=0.9\textwidth,trim=0 0 0 0,clip]{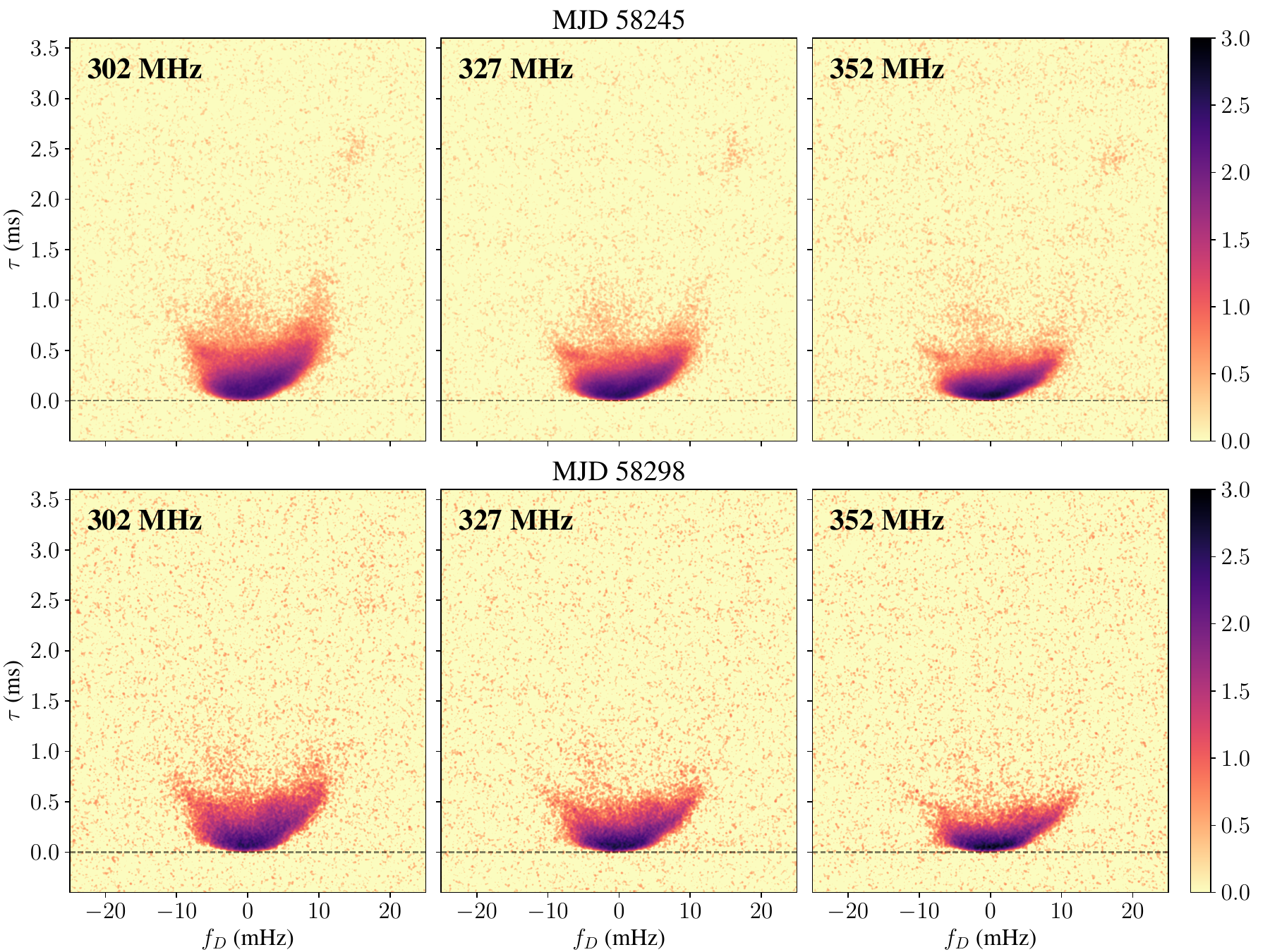}
  \caption{
    Wavefield solutions for three specific bands (of width $3.125$ MHz), for both our observations.
    The color scale corresponds to $\log_{10} |W(\tau, f_D)|^2$ in arbitrary units.
    One sees that most power is contained within $\tau\lesssim0.8{\rm\,ms}$, but that the IRF extends to larger delays, with a faint structure at $\sim\!2.5{\rm\,ms}$. \vspace{5pt}
  }
  \label{fig:wf}
\end{figure*}

\begin{figure}
  \centering
  \includegraphics[width=0.45\textwidth,trim=0 0 0 0,clip]{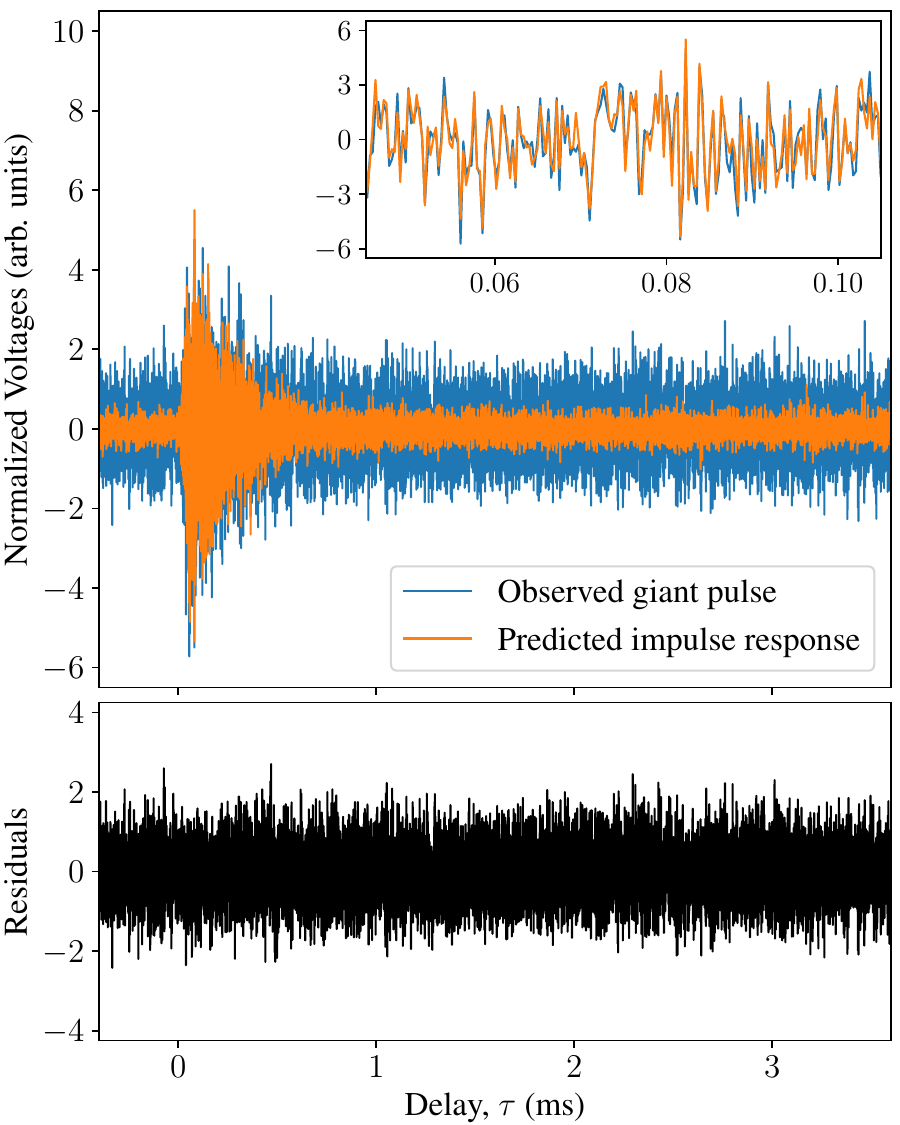}
  \caption{
    Comparison between the predicted impulse response and an observed giant pulse at $327$ MHz.
    As can be seen from the zoom-in as well as from the flat residuals, there is good correspondence between the data and the model.
  }
  \label{fig:pred}
\end{figure}

For our solutions of the wavefield, we restrict $\fD$ to the range $(-25, +25) \, \mathrm{mHz}$, with $N_{\fD} = 1125$ equally-spaced points.
Since the scintillation timescale is around $70 \, \mathrm{s}$, we do not expect to see signals at higher differential Doppler shifts.
The number of components is larger than required by lengths of our data sets (especially for the shorter observation), but given our choice of OMP for sparse approximation, this makes no difference to the result.

For the warm start described in Section~\ref{ss:amplitude}, we use a stopping criterion of $\gamma = 12$, to ensure that the approximate wavefield solution is virtually guaranteed to not inclue any noise components. This avoids biases when estimating the amplitudes of individual giant pulses. Furthermore, we restrict the warm start solution to a delay range of $0 \le \tau < 1.44{\rm\,ms}$, i.e., we exclude the
high-$\tau$ portion which contains very little power and noise dominates.

With the amplitudes in hand, we then construct the final wavefield, extending out to a delay range of $-0.4 \le \tau < 3.6{\rm\,ms}$ and using a more relaxed stopping criterion of $\gamma = 5$. This value of $\gamma$ is chosen to ensure that we can see even faint features at large $\tau$, but comes at the cost of allowing a larger fraction of noise into the solution.

After the removal of relative time offsets between polarizations (as described in Section~\ref{sec:obs}), we can expect both polarizations to have almost the same impulse response (this assumption is tested in Section~\ref{ssec:polval}). We ignore the higher-order effects of Faraday rotation which are insignificant for our observation given the source's low RM.

We can, therefore, make wavefields which are polarization independent by treating each polarization as a separate measurement of the same IRF.
In Figure~\ref{fig:wf}, we show these wavefield solutions for both observations across multiple bands.
In Figure~\ref{fig:pred}, we compare the predicted IRF from the modeled wavefield with the observed signal from a bright giant pulse.
In the zoomed inset of the top panel, we can see the strong correspondence between the predicted IRF and the observed giant pulse signal.
However, this correspondence seems stronger than expected given the amount of noise in the observed signal.
We further investigate this overfitting in Section~\ref{sec:validation}.

\begin{figure*}
    \centering
    \includegraphics[width=0.75\textwidth,trim=0 0 0 0,clip]{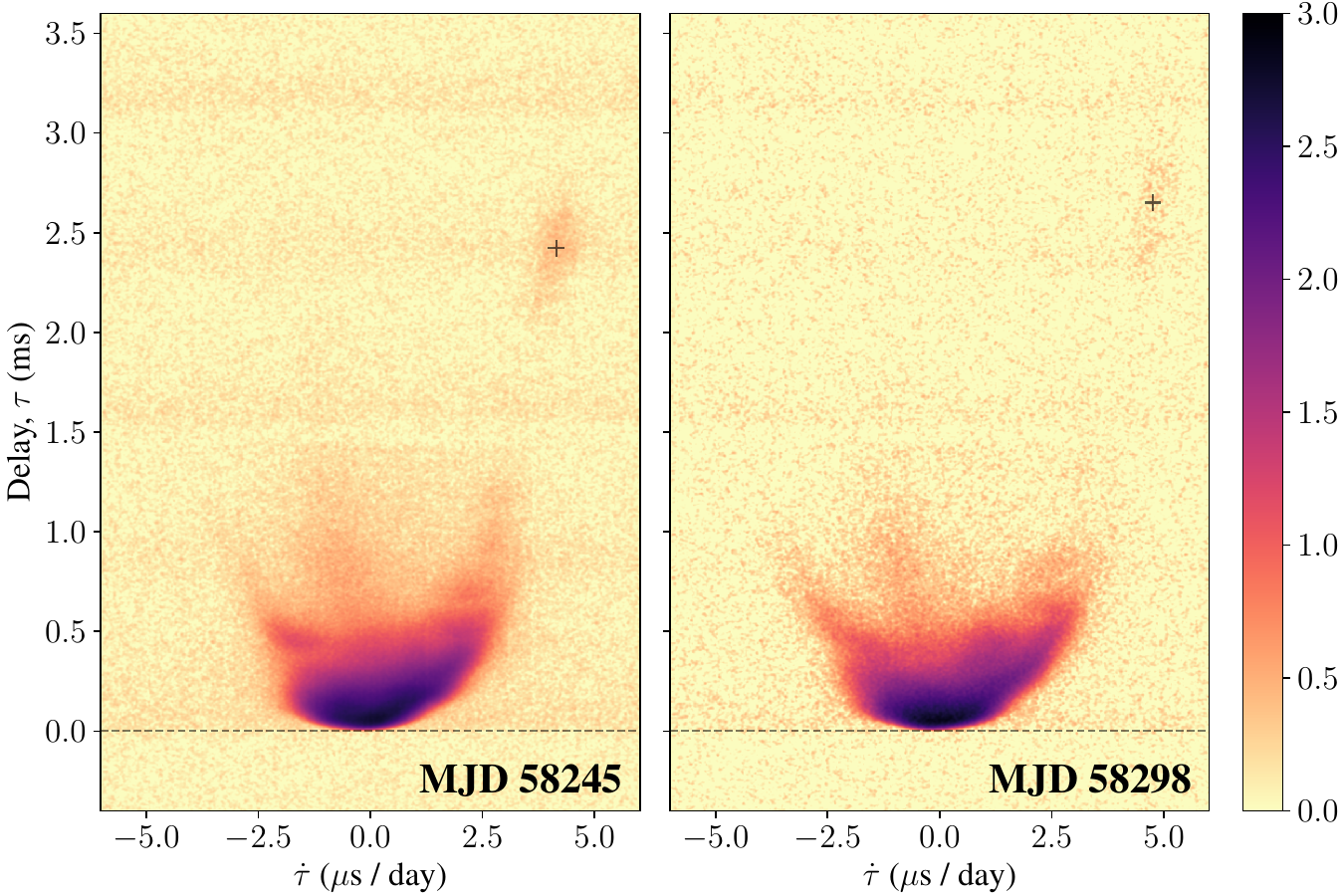}
    \caption{The total intensity of the wavefields in the two observations (with the wavefield intensities of the 19 subbands aligned using $(\tau, \dot{\tau})$ coordinates before summing).
    The color scale corresponds to $\log_{10} |W(\tau, \dot{\tau})|^2$ in arbitrary units.
    A cross denotes the measured center of the $2.4$ ms feature in both panels.
    The excess noise around $\tau=0$, 1.6, and $3.2{\rm\,ms}$ (as well as, more faintly, at 0.8 and $2.4{\rm\,ms}$), is due to the regular pulse emission.
    An interactive version of this figure, which shows the $(\tau, \dot{\tau})$ wavefields for each frequency band, is available at \url{https://www.astro.utoronto.ca/~mahajan/interactive/interactive_wavefield.html}. \vspace{10pt}}
    \label{fig:combined}
\end{figure*}

We can see that the primary structure in the wavefield has a characteristic parabolic envelope which is often seen in other observations of pulsar scintillation \citep{Walker2008, Stinebring2022, Main2023}. We also see no signal at negative delays, which is expected since the IRF must be causal. The structures in the wavefields appear to get wider in $\fD$ with increasing frequency. This is due to the frequency scaling of $\fD$. We can, instead, transform the wavefields to a more natural $(\tau, \dot{\tau})$ space via the transformation $\fD = \dot{\tau} \nu$. In this space, common features across frequency bands are aligned in $(\tau, \dot{\tau})$. As a result, we can get a higher signal-to-noise wideband representation of the total intensity in the wavefield via $\sum_\nu \left| W(\tau, \dot{\tau}) \right|^2$.
Figure~\ref{fig:combined} shows this combined $(\tau, \dot{\tau})$ wavefield for both observations. An interactive version of the figure also shows $\left| W(\tau, \dot{\tau}) \right|^2$ across the frequency bands, where the correspondence of features across frequency is visually evident.

\section{2.4 ms feature} \label{sec:feature}

In the wavefield for MJD 58245, a relatively bright feature can be seen at high delay, of around $2.4$ ms (which we refer to as the ``2.4 ms feature'' from here on), approximately $1.5$ times the rotational period of the pulsar.
A corresponding, but fainter, feature at a slightly higher delay can be seen on MJD 58298.
We fit the features with a bivariate (2D) Gaussian in the $(\tau, \dot{\tau})$ space to measure the location of the feature's center, which is denoted by a cross in Figure~\ref{fig:combined}. The $\tau$ and $\dot{\tau}$ values at the center of the Gaussians that we fit are provided in Table~\ref{table:1}.
\begin{deluxetable}{ccc}
\setlength{\tabcolsep}{15pt}
\tablecaption{2.4 ms feature center measurements \label{table:1}}
\tablenum{1}
\tablehead{\colhead{Time} & \colhead{Delay, $\tau$} & \colhead{$\dot{\tau}$} \\
\colhead{(MJD)} & \colhead{(ms)} & \colhead{($\mu$s / day)} }
\startdata
58245.38 &  2.422 $\pm$ 0.004 &  4.15 $\pm$ 0.01 \\
58298.24 &  2.650 $\pm$ 0.016 &  4.75 $\pm$ 0.02 \\
\enddata
\tablecomments{The MJD times listed are at the center of our observations.}
\end{deluxetable}

Assuming that $\dot{\tau}$ changes linearly between the two observations, one expects $\Delta\tau = \langle\dot\tau\rangle\Delta t=0.235 \, {\rm ms}$, in agreement with our measured change in delay, $\Delta\tau = 0.228 \pm 0.016 \, {\rm ms}$.
This implies that the features in both observations most likely correspond to the same scattering structure.
On the sky, this structure is moving away from the line of sight to the pulsar, leading to a higher delay in the later observation.

\begin{figure}
  \centering
  \includegraphics[width=0.45\textwidth,trim=0 0 0 0,clip]{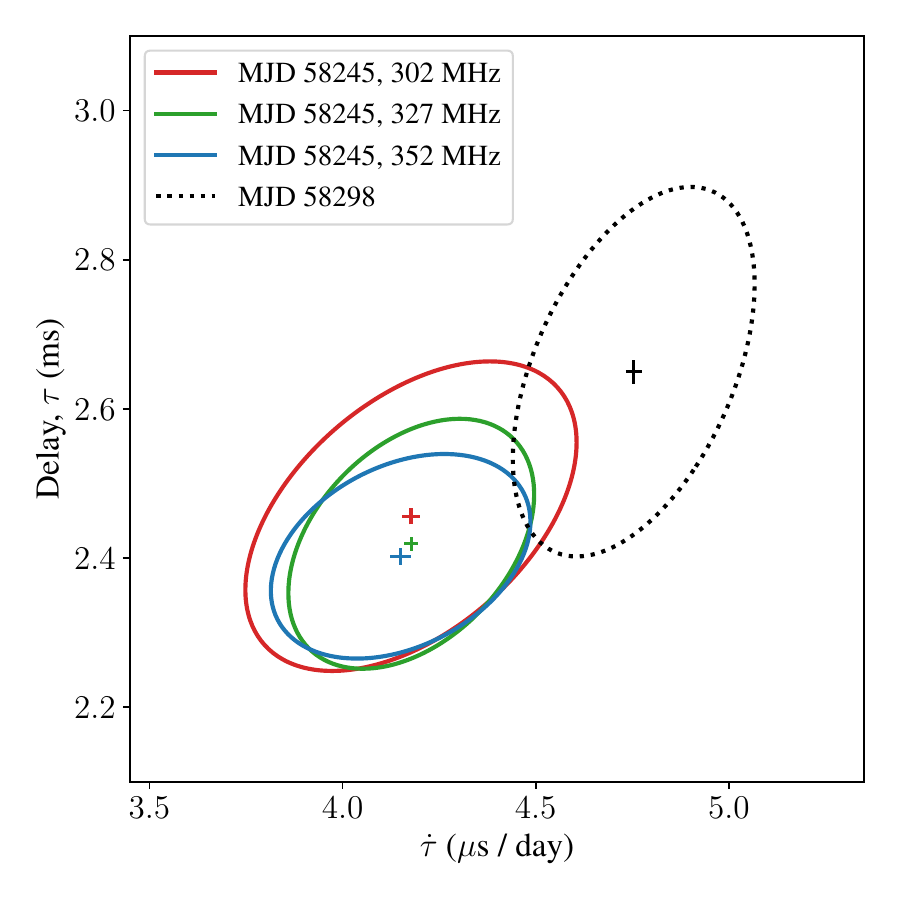}
  \caption{
    Best-fit $\tau$ and $\dot\tau$ (with error bars) inferred from Gaussian fits to the 2.4 ms feature across three frequency bands on MJD 58245.
    The contours mark the $1 \sigma$ widths of the Gaussians.
    The center and widths for a fit for the total intensity on MJD 58298 are shown for comparison.
  }
  \label{fig:feature}
\end{figure}

We also investigate the frequency dependence of the delay for the 2.4 ms feature by fitting a 2D Gaussian to the feature on a rolling average of 3 frequency bands.
In Figure~\ref{fig:feature}, we show the $1 \sigma$ contours of these 2D Gaussian fits for different frequencies on MJD 58245.
Following \cite{Brisken2010}, who fit the angular offsets $\theta$ from the line of sight to a power law of the form $\theta\propto\lambda^\gamma$, we try fitting to $\tau\propto \nu^{-2\gamma}$ (using that for a single, thin screen, $\tau\propto\theta^2$).
Using the centers of the 2D Gaussian fits for different frequency bands, we measure $\gamma = 0.061 \pm 0.005$.
This value of $\gamma$ is similar to the value measured by \cite{Brisken2010} for their ``1 ms feature''.
According to \cite{Simard2018}, this value for $\gamma$ is plausible for features caused by lensing from an under-dense refractive plasma sheet.
It should be noted that the feature does not merely shift downwards in $\tau$ with increasing frequency, but also get smaller in size.
Thus the apparent shift in $\tau$ is could simply be a consequence of high-delay paths becoming less favorable at higher frequencies.
We do not find a significant frequency dependence along $\dot{\tau}$.
\section{Validation}
\label{sec:validation}

\begin{figure*}
  \centering
  \includegraphics[width=0.9\textwidth,trim=0 0 0 0,clip]{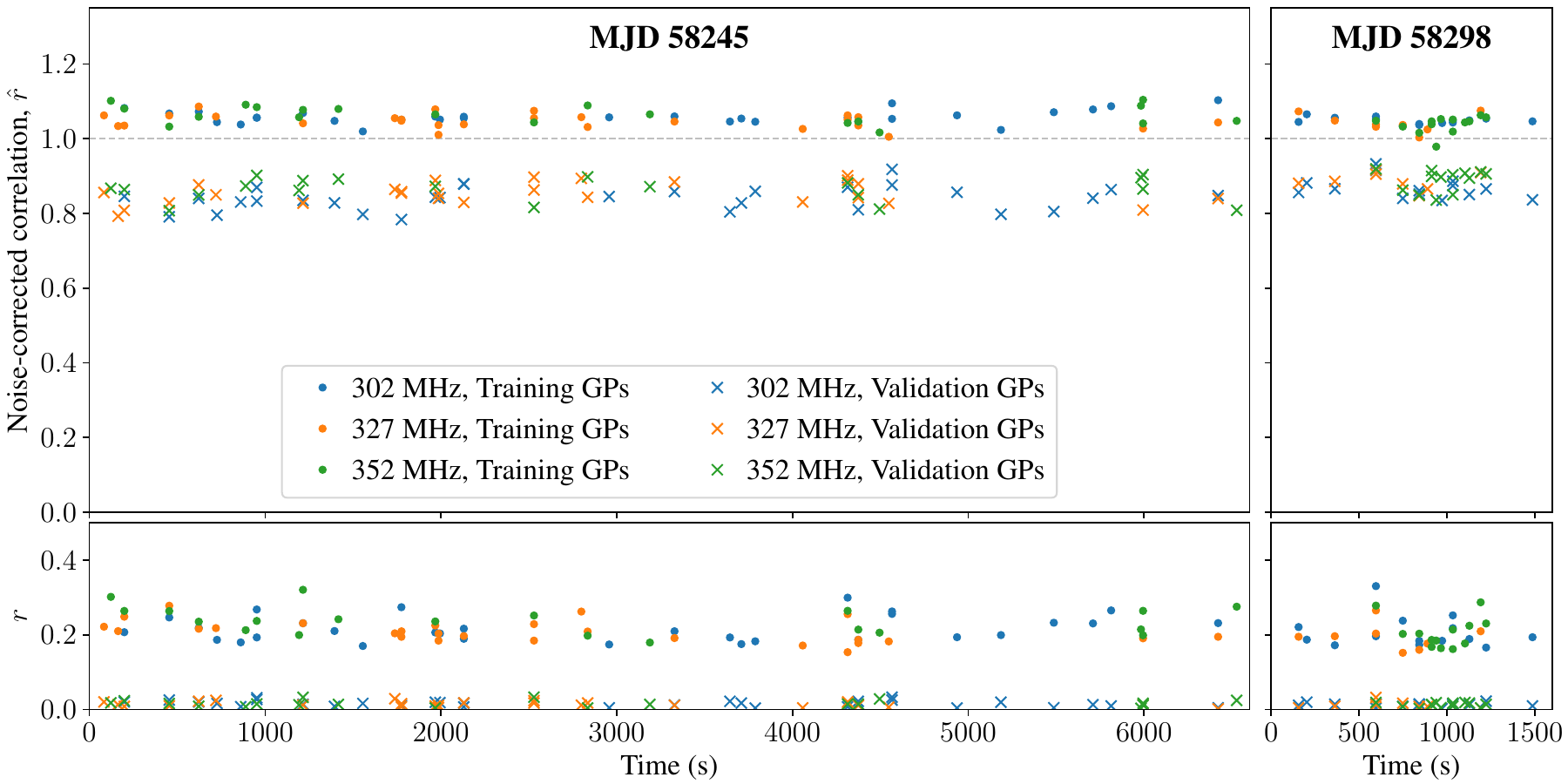}
  \caption{
    \textit{Top:} Noise-corrected normalized correlations between giant pulse data and predicted IRFs for $0 \leq \tau \leq 0.8$ ms, where most of the signal lies.
    The correlations are corrected for contributions by noise in both data and model, but not for correlation in the noise between the two.
    \textit{Bottom:} Normalized correlations for the pure-noise regions of $\tau < 0$ and $\tau > 3{\rm\,ms}$ between data and model. The correlation between the model and the training set confirms that they share correlated noise, while the lack of correlation with the validation set shows that there are no systematic effects shared between all pulses. \vspace{5pt}
  }
  \label{fig:valid}
\end{figure*}

In order to measure the performance of the modeled wavefields, we compare how well the predicted impulse responses correlate with the giant pulse data. Given two signal vectors, $\vect{x}$ and $\vect{y}$, we can compute a normalized correlation
\begin{equation}
r = \frac{\left| \vect{x}^* \, \vect{y} \right|}{\|\vect{x}\| \|\vect{y}\|},
\end{equation}
which takes on values in the range $[0, 1]$. However, if we have two noisy vectors, $\hat{\vect{x}}$ and $\hat{\vect{y}}$, which include additional Gaussian noise terms, then we compute a noise-corrected normalized correlation
\begin{equation}\label{eq:corr}
\hat{r} = \frac{\left| \hat{\vect{x}}^* \, \hat{\vect{y}} \right|}{\sqrt{\left( \|\hat{\vect{x}}\|^2 - N {\sigma_x}^2 \right) \left( \|\hat{\vect{y}}\|^2 - N {\sigma_y}^2 \right)}},
\end{equation}
where $N$ is the length of the vectors, and ${\sigma_x}^2$ and ${\sigma_y}^2$ are the respective variances of the noise terms in the vectors. If the noise terms are uncorrelated, then $\hat{r} \approx r$. However, if the noise terms are correlated, the numerator in Equation~\ref{eq:corr} will be inflated by an additive term, $N \sigma_{xy}$ where $\sigma_{xy}$ is the covariance between the two respective noise terms.

\subsection{Cross-Validation}

We employ cross-validation with 13 ``folds'', i.e., we divide the giant pulses into 13 sets, each distributed similarly in time and polarization.
We take out one of these sets as a ``validation set'', and solve for an unpolarized wavefield using the remaining giant pulses, called the ``training set''.
We do this 13 times such that every giant pulse is in a validation set once.

In Figure~\ref{fig:valid}, we show the noise-corrected correlations between bright giant pulses and the predicted IRFs (over $0 \le \tau \le 0.8$ ms).
One sees a clear difference between the validation and training sets, with the noise-corrected correlation for the validation set at $\hat{r} \simeq 0.86$, while that for the training set is at $\hat{r} \simeq 1.05$.
The inflation in $\hat{r}$ for the training set is due to correlated noise, induced by overfitting: with our choice of a relatively low value of $\gamma$, some of the noise in the data has been included in the model.
That this is indeed the cause of the difference can be seen in the bottom panel of Figure~\ref{fig:valid}, where we show the normalized correlation using only the pure-noise regions.
As expected, the noise of pulses in the validation set is uncorrelated with that of the predicted IRFs, while the noise of pulses in the training set is slightly correlated.

It should be noted that the mean $\hat{r}$ for the validation set is slightly higher for MJD 58298 compared to MJD 58245, despite wavefield solutions using the identical parameters. This is likely due to the higher overall quality of the solved wavefields from MJD 58298 where we find a higher detection rate of giant pulses. The quantity of $\hat{r}^2 \simeq 0.75$ has a convenient interpretation as being the fraction of total integrated intensity of the true impulse response that is captured by the modeled wavefields.

\subsection{Comparing polarizations}
\label{ssec:polval}

In making polarization-independent wavefields, we make the assumption that the impulse response is the same between polarizations.
Since Faraday rotation effects cause circular birefringence as light propagates through magnetized ISM, we know this assumption can be false in general.
In practice, one could measure the average RM from bright giant pulses and correct for it properly with an inverse filter.
In our case, however, the RM is small enough that simply correcting for a constant time delay (as described in Section~\ref{sec:obs}) between polarizations is sufficient, as the higher-order effects are negligible in our narrow, $3.125{\rm\,MHz}$ bands.

In order to test this assumption, we can measure the correlation between the two polarizations.
For individual giant pulses, we know that the noise terms are correlated between polarizations due to the underlying regular pulsar emission and possibly due to imperfections in the receiver (for example, the polarizations may not be perfectly orthogonal).
Instead, we estimate the IRF for giant pulses within $20$s blocks (using a rank-1 approximation as described in Equation~\ref{eq:svd}).
This estimate should have uncorrelated noise across the two polarizations, but the IRF itself should stay relatively constant over a $20$s period. We measure a mean noise-corrected correlation between the the left- and right-circular polarizations of $\hat{r} \simeq 0.98$.
Thus, we find that the two polarizations seem to largely contain the same information about the IRF.

\subsection{Predictions in Gaps}

We also experiment with how our technique performs in regions where there are no data.
We do this by solving the wavefield on a subset of giant pulses where we exclude data from a small span of time in the middle of the observation.
We find that for any span of time larger than the scintillation timescale, the predicted IRF fails to correlate with the data in the gap, with the noise-corrected correlation dropping to zero for the entirety of the gap. Even with a short $1$ minute gap, the noise-corrected correlation at the center of the gap drops to $\hat{r} \simeq  0.5$. We see the same effect when solving the wavefield on only the central $1$ hour of data on MJD 58245. The correlation drops to zero almost immediately outside the bounds of where data were available.

This implies that, for our data set at least, every ``scintillation timescale''-sized period of time contains unique and non-redundant information about the time-varying impulse response, and that the predicted impulse response cannot be trivially interpolated into regions where there are no data, if the gaps are of order $t_\mathrm{scint}$ or longer.

\section{Intrinsic Pulsar Emission}
\label{sec:intrinsic}

\begin{figure*}
  \centering
  \includegraphics[width=0.7\textwidth,trim=0 0 0 0,clip]{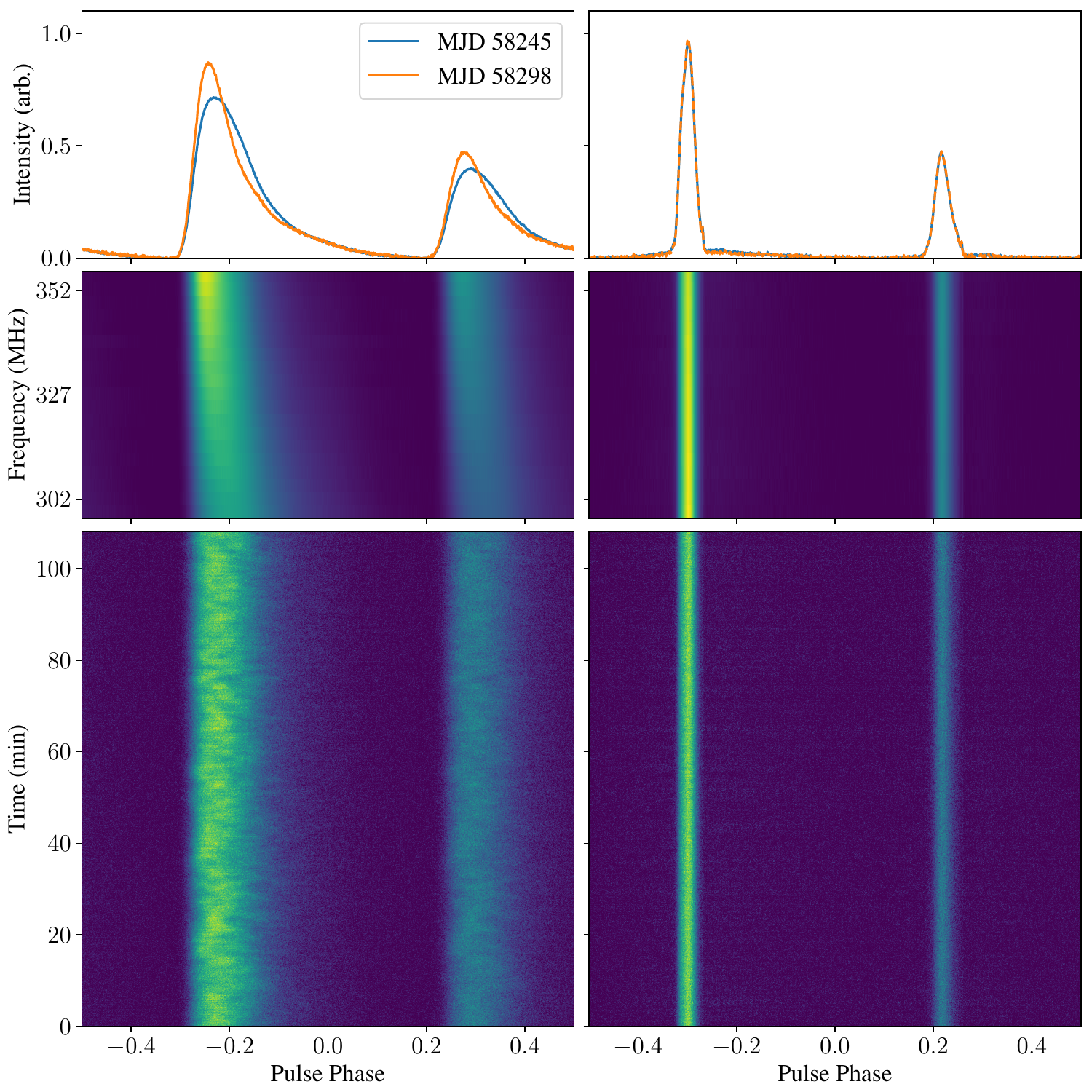}
  \caption{
    A comparison between the observed pulse profile (left panels) and the recovered ``intrinsic'' pulse profile (right panels) for PSR B1937+21.
    \textit{Top:} The average pulse profiles at 327 MHz (bandwidth of 3.125 MHz) for our two observations, at MJD 58245 and MJD 58298.
    On the right, the profile for MJD 58298 is plotted with a dashed line to help visually identify both profiles.
    \textit{Middle:} The average pulse profiles for the observation on MJD 58245 across the 19 frequency bands.
    \textit{Bottom:} The pulse profiles at 327 MHz (bandwidth of 3.125 MHz) across time, for the observation on MJD 58245.
  }
  \label{fig:profile_freq}
\end{figure*}

\begin{figure*}
  \centering
  \includegraphics[width=0.8\textwidth,trim=0 0 0 0,clip]{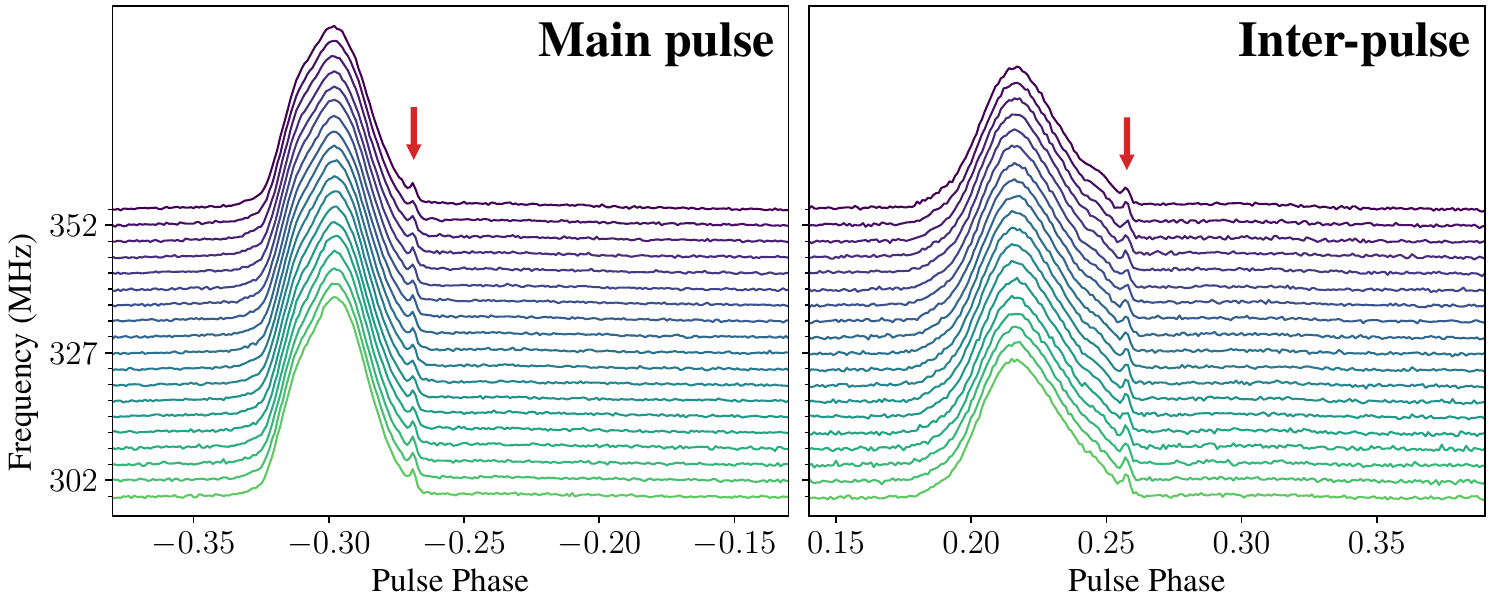}
  \caption{
    A closer look at the main pulse (left) and inter-pulse components (right) in the recovered ``intrinsic'' pulse profile across 19 frequency bands.
    Red arrows point at a bump caused by giant pulse emission.
    The vertical axis for each profile is in intensity with arbitrary units.
    Scaling is different between the left and right panels to help see details; the inter-pulse component is approximately half as bright at its peak as the main pulse component.
  }
  \label{fig:profile_comp}
\end{figure*}

The intrinsic pulsar emission can be recovered via deconvolution. We observe a signal $y = (h * x) + \epsilon$, and our goal is to compute a filter, $g$, such that $\hat{x} = g * y$ approximates $x$ well. We define a regularized inverse filter,
\begin{equation}
    G(\nu) = \frac{H^*(\nu)}{|H|^2 + \mu}
\end{equation}
where $H$ and $G$ are the Fourier transforms of $h$ and $g$ respectively, and $\mu$ is a regularization parameter. When $\mu = 0$, then $g$ is simply the inverse filter of the IRF $h$.
However, since direct inverse filtering can be unstable due to division by values close to zero, a small positive $\mu$ can be used as a form of regularization.
Since we do not know the true IRF but only a noisy approximation of $h$, this deconvolution scheme is not optimal.
We use a constant value of $\mu = 0.01$ which is determined by trial-and-error -- higher values of $\mu$ result in undesirable artifacts in the recovered signal, while lower values result in a lower signal-to-noise ratio in the recovered signal.

We apply the regularized inverse-filter, $g(\tau)$, to the observed signal, $y(t)$, to recover the ``intrinsic'' emission signal, $\hat{x}(t)$.
Using pulsar timing models from NANOGrav \citep{NanoGrav}, we fold $\hat{x}(t)$ to get an ``intrinsic'' pulse profile in the same way that the observed signal is usually folded.
Figure~\ref{fig:profile_freq} shows a comparison between the observed and recovered intrinsic pulse profiles.
For both observations, the intrinsic profiles look almost identical despite the observed profiles being noticeably different due to changes in scattering between observations.
There is also a strong correspondence in the intrinsic profile across the 19 frequency bands (also shown in Figure~\ref{fig:profile_comp}).
In our work so far, all frequency bands have been processed entirely independently of each other.
Thus, this correspondence provides additional evidence for the validity of our modeled wavefields and the predicted impulse response.
In Figure~\ref{fig:profile_freq}, one also sees that the intrinsic profile is stable across our 2 hour observation, even though the observed profile varies significantly due to scintillation effects.
Since we assumed earlier that the impulse response has a constant total power over time, the total intensity of the recovered pulse profile is likely inaccurate over short timescales.
In principle, the observed pulse profile could be used to calibrate the variations of the total IRF power over time, which could be used to adjust the amplitudes determined in Equation~\ref{eq:ampsolve}.

In Figure~\ref{fig:profile_comp}, we find a narrow bump on the trailing shoulder of both pulse components which we identify as the giant pulse emission component.
Almost all of the detected ``impulses'' from our giant pulse search come from these two regions in pulse phase.
We defer the in-depth analysis of the pulse phase distribution of detected pulses and the intrinsic pulsar emission to a future publication, which is in preparation.

\section{Discussion} \label{sec:discussion}

\subsection{Solved Wavefields}

In Figure~\ref{fig:combined}, we can see faint but periodic bands in the background noise in delay. This is a manifestation of the regular pulse emission underlying the giant pulse data.
In our sparse approximation technique, we use the same stopping criterion for all delays.
Thus, at delays where the regular pulse emission contributes, we see slightly brighter noise bands.

The wavefield we recover is relatively dense and filled in within its envelope.
We see at least three main ``arms'' localized in $\dot{\tau}$ extending up into higher delays, in both observations.
The wavefield recovered via cyclic spectroscopy by \cite{Walker2013} for observations of PSR B1937+21 at 428 MHz also show a relatively filled-in primary wavefield structure, although it extends only out to about $\sim\!0.5$ ms, as expected given the higher observing frequency.

In constrast, the wavefields recovered by \cite{Walker2008} and \cite{Baker2022} for observations of PSR B0834+06 around 320 MHz show thin parabolic structures which indicates highly localized, anisotropic scattering.
It is possible that the wavefield structure in PSR B1937+21 is caused by the signal propagating through several scattering structures or screens.
Since PSR B1937+21 lies in the Galactic plane ($b = -0\fdg3$) at a distance of $2.9{\rm\,kpc}$ \citep{Ding2023}, while PSR~B0834+06 is at higher Galactic latitude ($b=26\fdg3$) and is at only $0.65{\rm\,kpc}$, it is not implausible that there are more scattering structures along the line of sight to PSR~B1937+21.

Another piece of evidence pointing to multiple scattering structures is that the bright 2.4 ms feature is quite extended in both $\tau$ and $\dot{\tau}$.
If this was an independent scattering structure in the ISM, it would have to be quite large, with many scattering points.
It seems more plausible that instead the situation is similar to what is inferred for the ``$1{\rm\,ms}$'' feature in PSR~B0834+06 \citep{Liu+2016,Zhu2022}, viz., that the large $\tau$ and $\dot\tau$ reflect scattering off a few strong scattering points quite far from the line of sight, and that the extent represents further scattering also by other structures closer to the line of sight.

\subsection{Comparison with other techniques}

All other methods of recovering wavefields so far operate on dynamic spectra.
A conventional dynamic spectrum is created by folding channelized intensity data into a pulse profile across many sub-integrations.
Since this removes phase information, it makes reconstructing the impulse response of the ISM much more difficult.
Furthermore, since usually several phase bins are necessary to sufficiently localize the signal in pulse phase, the spectral resolution is inherently limited by the pulse width, and generally it is impossible to determine the impulse response beyond the pulse period.
Hence, a dynamic spectrum would not contain information on features like that at $2.4{\rm\,ms}$ we detect, since that occurs at a delay of 1.5 times the rotation period of PSR B1937+21.

The situation is better with cyclic spectroscopy, as it keeps most of the phase information \citep{Walker2013}.
On the other hand, it relies on the assumption that the pulsar signal is cyclostationary, and thus suppresses information in the observed signal which does not conform to the cyclic frequency provided.
For this reason, the recovered intrinsic pulse profiles of PSR~B1937+21 of \cite{Walker2013} do not exhibit a giant pulse emission bump.


\subsection{Wavefield Performance}

We find that our solved wavefields fail to interpolate across even relatively short gaps of a minute or so in the data without significant loss in performance, implying that the time-varying impulse response contains unique non-redundant information across every scintillation time.
This reflects the fact that the wavefield is dense, with an extent in $f_D$ of roughly the inverse of the scintillation time.
This is easiest to understand if there are multiple scattering structures along the line of sight to the source.

This does not exclude that a sparser wavefield would have greater interpolation performance, since there would be less overall information to encode, and every short period of time would contain largely redundant information.
Such a wavefield might be found for a different pulsar with a less complicated scattering geometry along the line of sight, or perhaps for PSR B1937+21 at higher frequencies, where scattering is less efficient.

In principle, it may also be that even at our frequency it is possible to resolve the scattering points by including larger frequency chunks, if the scattering geometry does not vary strongly with frequency and is close to being resolved in our present bands.
Modest evolution with frequency is suggested by our measurements of the $2.4{\rm\,ms}$ feature, as well as from observations of PSR B0834+06 which show similar structures in neighboring frequency bands \citep{Brisken2010}.
We have not pursued this with our data, both since we worry about complications arising from the gaps in frequency coverage caused by the polyphase filter and because it would seem better to start at somewhat higher frequency where the situation should be less complex.
We note, though, that with wider frequency bands it may be required to account for the evolution of structures with frequency, perhaps using more physically motivated models such as those of \cite{Simard2018}.

Finally, we note that features in the wavefield are expected to move in delay (given non-zero $\dot{\tau}$) even across observations a few hours long. Thus, we suspect that a solving the wavefield over a wavelet basis, localizing features in time, rather than a Fourier basis may improve performance by promoting sparsity.
\section{Conclusions and Ramifications} \label{sec:ramifications}

In this paper, we present a novel technique for giant pulse search by pattern matching the raw voltage signals to find pulses at a higher rate than conventional methods.
We have made available the baseband snippets for all 13,025 giant pulses found using this technique as a dataset on Zenodo at \dataset[doi:10.5281/zenodo.7901384]{\doi{10.5281/zenodo.7901384}}.

This technique can be applied to study giant pulse emission in other pulsars, such as PSR B1957+20 \citep{Main2017} and PSR J1823-3021A \citep{Abbate2020}, both of which are known to emit giant pulses at a high rate.
It might also be used in combination with cyclic spectroscopy, to give a better starting point for inferring the impulse response from the cyclic spectra, and/or to verify that that the resulting responses properly de-scatter giant pulses.
Another application would be to attempt recovering weaker bursts in repeating fast radio bursts such as FRB20201124A which show similar scintillation patterns for bursts nearby in time \citep{Main2022}.

A basic assumption of our method is that giant pulses share the same impulse response.
Hence, by using it to test whether this is the case, one can determine if a giant pulse emission region is being spatially resolved by scattering, as has been inferred for the Crab Pulsar \citep{Lin2023}. 
The solved wavefields themselves can be a useful tool to study the structure of the ISM.
For instance, comparing wavefields of different circular polarizations, one can study magnetic fields along the line of sight to the source.

As we demonstrated, the intrinsic emission of a pulsar can be recovered via deconvolution using the predicted IRFs.
We will present an analysis of the intrinsic emission for PSR B1937+21 in an upcoming paper.
This technique could be even more useful at lower frequencies where the emission profile of pulsars is often so dominated by stronger scattering effects that the pulse profile is completely washed out \citep{Kondratiev2016}.
The ability to recover the intrinsic emission profile at frequencies as low as 100\,MHz would aid our understanding of the radio emission mechanism in pulsars.

\begin{acknowledgements}
  We thank the Signal Processing community on Stack Exchange for many helpful insights, and the Toronto scintillometry group, in particular Ue-Li Pen and Rebecca Lin, for discussions.
  This research has made use of NASA's Astrophysics Data System Bibliographic Services.
  Computations were performed on the Niagara supercomputer at the SciNet HPC Consortium \citep{Loken2010, Ponce2019}.
  SciNet is funded by: the Canada Foundation for Innovation; the Government of Ontario; Ontario Research Fund - Research Excellence; and the University of Toronto.
  MHvK is supported by the Natural Sciences and Engineering Research Council of Canada (NSERC) via discovery and accelerator grants, and by a Killam Fellowship.
\end{acknowledgements}

\facility{Arecibo (327 MHz Gregorian). The data used in this publication was obtained as part of observing project P3229.}

\software{
    Pulsarbat \citep{pulsarbat},
    Baseband \citep{baseband},
    Numpy \citep{numpy},
    Astropy \citep{astropy:2013, astropy:2018, astropy:2022},
    Scipy \citep{scipy},
    Matplotlib \citep{matplotlib},
    Dask \citep{dask},
    TEMPO2 \citep{tempo2:1, tempo2:2}
}

\appendix
\section{Impulse Estimation} \label{app:impulse}

Given a noisy complex-valued signal consisting of an impulse, we wish to find the best estimate of the location of the impulse in time.
Essentially, given $z(t) = a \, \delta(t - t_0) + \epsilon$, we want to estimate $t_0$.
Here, the noise term, $\epsilon$, is assumed to be additive Gaussian white noise.
The Fourier transform of $z(t)$ would be a noisy signal consisting of a single complex sinusoid, the frequency of which corresponds directly to $t_0$. Thus, the problem of locating an impulse in time is the same as that of measuring the tone of a single-frequency signal.
This problem has been studied extensively in the engineering literature, and much of the work below draws heavily from proofs presented in \cite{Rife1974}.

In practice, we observe a band-limited discrete-time version of $z(t)$. The discrete Fourier transform (DFT) of this signal is given by,
\begin{equation}
    Z_n = a \exp\left(\frac{-2\pi i t_0 n}{N}\right) + \epsilon,
\end{equation}
where $a$ is the complex amplitude of the impulse, and $\epsilon \sim \mathcal{CN}(0, N \sigma^2)$ is the noise term ($\mathcal{CN}$ refers to the complex-valued normal distribution). The signal-to-noise ratio of the impulse is given by $S/N = |a|^2 / \sigma^2$. The factor of $N$ in the noise variance is due to the convention in normalizing DFTs. In time-domain, this translates to our assumption of additive Gaussian white noise with variance $\sigma^2$.

Intuitively, we can see that by taking the discrete-frequency, continuous-time inverse Fourier transform of $\vect{Z}$,
\begin{equation}
    f(t) = \frac{1}{N} \sum_{n} Z_n \exp\left(\frac{2 \pi i t n}{N}\right)
\end{equation}
we can get a maximum-likelihood unbiased estimator of $t_0$,
\begin{equation}
    \hat{t} = \underset{t}{\operatorname{argmax}} \left|f(t) \right|^2
\end{equation}
It can also be seen that $\hat{a} = f(\hat{t})$ is a maximum-likelihood estimator for the complex-amplitude of the impulse, $a$.

The variance of $\hat{t}$ is given by
\begin{equation}
    \operatorname{var}(\hat{t}) = \frac{3 }{2 \pi^2} \frac{\sigma^2}{|a|^2} \frac{N^2}{N^2 - 1} \approx \frac{0.152}{S/N}
\end{equation}
which is also the Cramér-Rao lower bound.

From this, we can see that the location for an impulse with $S/N = 18$ can be estimated with an error of $\sigma_{t} \sim 0.09$ samples (which is $30$ ns when using a sample spacing of $320$ ns, as is the case in our data).

\bibliographystyle{aasjournal}
\bibliography{thesis}

\end{document}